# Deep-Fill: Deep Learning Based Sinogram Domain Gap Filling in Positron Emission Tomography

**Isaac Shiri[1], Peyman Sheikhzadeh*[1,2], Mohammad Reza Ay [1,3]**

1. Research Center for Molecular and Cellular Imaging, Tehran University of Medical Sciences, Tehran, Iran
2. Department of Nuclear Medicine, Vali-Asr Hospital, Tehran University of Medical Sciences, Tehran, Iran.
3. Department of Medical Physics and Biomedical Engineering, Tehran University of Medical Sciences, Tehran, Iran

**Abstract:**

One of the major challenges in design and developing of PET scanners is the presence of inactive areas between the detector blocks which degrade the image spatial resolution and leads to streaking artifacts especially when we employ analytical image reconstruction. The aim of this study is to assess the feasibility of generating the gap-free PET image using the deep convolutional encoder-decoder in sinogram space. The gap-corrupted sinograms of simulated HRRT scanner, sinograms without gaps as ideal/ground truth and predicted sinograms owing to the implemented our deep-fill method were quantitatively compared. In total, 1293 phantom images divided into three main sets of training 1000, 150 Validation, and 143 test set. The 1000 training images were augmented using affine transformations with various sub-transforms including rotation (rotate randomly), translation to 12000 Image with 6 frequencies of 2 main methods. The Deep-Fill architecture consists of an encoder and a decoder part, and it is composed of convolution operation, max pooling, ReLU activation, concatenation, and up convolution layers. The gap image is going through the network along with all possible paths then the gap-free image was generated by decoder part of the network. The quality of the generated images was quantitatively assessed by different quality metrics in both sinogram space and reconstruction images. We demonstrated that deep learning–based approaches applied to inter-detector gap filling can recover the missing data in sinogram with high quantitative accuracy and have the potential to significantly improve the reconstructed image quality and prevent degradation of PET image quantification.

**Introduction:**

One of the major challenges in design and developing of PET scanners is the presence of inactive areas between the detector blocks. This challenge is more significant for scanner designs of organ dedicated in which the scanner could not be manufactured as a full circular ring scanner. ECAT HRRT brain PET is a case in point which comprised of eight detector module. The octagonal configuration of HRRT with eight gaps leads to a loss of approximately 18% of the sinogram dataset (when 3-D data are rebinned to 2-D stack of sinograms) (1). Also the produced gap challenge is unavoidable in animal PET scanner with a small ring diameter(1). Missing data in sinogram degrade the image spatial resolution and leads to streaking artifacts especially when we employ analytical image reconstruction such as the 2D filtered back-projection (FBP) algorithm that require a complete data set (1).

Different techniques was proposed in literature for compensation of inter detector gaps in sinogram domain or frequency domain of sinogram (2). One of the proposed methods relies on various forms of interpolation approach. Truncated and windowed Sinc, Bilinear, Bicubic, Gaussian, model-based methods and the constrained Fourier space (CFS), or interpolation filtering are examples of the interpolation approach (3). These approaches work relatively well if the gaps are small but they are sensitive to local variation, and also oscillate severely at the end points of the data range(1, 4-6). Another approach based on statistical framework such as maximum likelihood expectation maximization (MLEM) or maximum a posteriori (MAP). Expectation maximization (EM) algorithms perform better than the interpolation methods but needs much longer time. Slow convergence rates and noisy images by increase of iteration number are this algorithm drawback.

Thanks to the data availability and GPU computational resources, machine learning algorithms has been applied in different area of medical images (7, 8). Deep structured learning is subclass of machine learning algorithm, based on learning data representations. In deep learning, a model learns directly from data such as image, text and video to perform a desired task. Different application of deep learning techniques proposed in medical images analysis such as classification, object detection, region of interest segmentation and image transformation (9). Four main algorithms of deep learning containing convoluting neural network (CNN), restricted Boltzmann machine (RBM), sparse coding and Auto-Encoder were developed for different application (10). Convolutional Encoder-Decoder (CED) consists of a paired encoder (for efficient image compression while searching robust and spatial invariant image features) and decoder (the reverse process of the encoder to reconstructed output) (11).

Recent studies applied the different algorithm of deep learning such as CED and generative adversarial networks in PET modality such as high resolution PET image generation (12), image quality improvement (13) and PET simultaneous attention correction and reconstruction(14). the Xu *et al* (15) used CED to provide 200x Low-dose PET reconstruction images. Wang *et al* (16) applied the 3D conditional generative adversarial networks (3D-CGAN) and CED for high-quality image estimation from low dose PET images. Han *et al* (17) generated pseudo CT images from T1-weighted MR images in brain region. Liu *et al* (18) generated the pseudo CT by using T1-weighted MR and CED to labels tissue of brain (air, bone and soft tissue) for PET AC. Recently

Shiri *et al (19)* successfully applied the CED on direct brain PET image attenuation correction by direct mapping of non-attenuation correction PET image to measured attenuation correction PET images. The main aim of this investigation is to assess the feasibility of generating the gap free PET image using the deep convolutional encoder decoder in sinogram space.

**Material and Methods:**

**Data simulation**

We simulated HRRT PET scanner using GATE (Geant4 Application for Tomographic Emission) software (20). The specifications of simulated model were used based on real scanner configurations. A cylindrical source phantom with 31-cm diameter and 26-cm height, which is entirely cover the scanner axial and trans-axial FOVs was simulated (7). The sinograms of simulated phantoms were generated with 288 angular views and 256 radial bins. We extracted the gap pattern (see in Figure 1) from the simulated phantom.

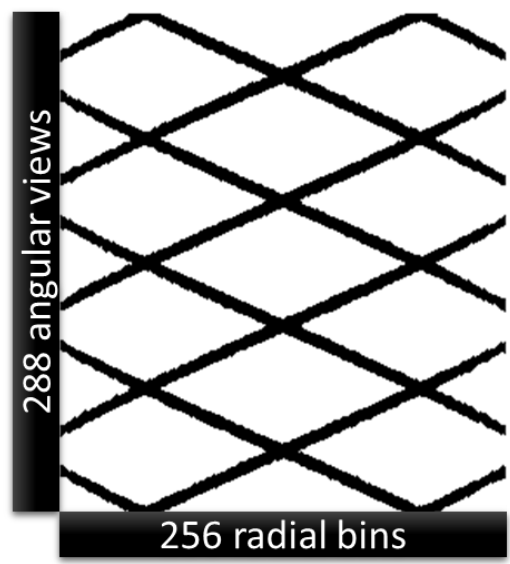

Figure1: The produced gap pattern from simulated HRRT scanner

For better assessment, we simulated ideal scanner without inter detector gaps. The ideal scanner provides us a gap-free sinograms which let us to comprise our final gap-corrupted sinogram with them and find absolute recovery ratio. Three sets of phantoms were constructed and imported to modeled HRRT scanner in order to accurately evaluate our techniques. The 3D Hoffman brain phantom inside HRRT scanner were also simulated (21). This phantom with a voxel size of 1.25

× 1.25 ×10 mm³, comprised of 19 slices and image matrix size of 196 ×196. The tracer was Fluorine-18 and the activity ratio of the grey matter: white matter: cerebrospinal fluid (CSF) was set to 5: 1: 0. Simulation were performed in 30 min scan time. For a second phantom, the voxelized XCAT phantom was used (22). The F-18 activity ratio of the grey matter: white matter: tumor was set to 4: 1. The non-attenuated phantom was used and GATE simulation with inserted XCAT phantom was performed in 300 s. Finally, Zubal head phantom as a voxelized brain source (23) were simulated. This phantom consists in a segmented MR image with voxel size of 1.09×1.09×1.4 mm³ with matrix size of 256 ×256×128. Some of important regions of the phantom based on raclopride C-11 uptake in human brain were considered for the simulations such as putamen and caudate nucleus. The sinogram resulting was reconstructed using in-house two dimensional FBP MATLAB code. The FBP algorithm is of linear behavior for image reconstruction thus preferred over the iterative algorithms.

**Data Augmentation**

To increase the amount of data in dataset, affine transformations with various sub-transforms including rotation (rotate randomly), translation (moving the image along the X or/and Y direction) were applied to data. Data augmentation help the network to prevent from data memorizing during training. In total the 1293 of image of phantom divided to three main set of training 1000, 150 Validation, and 143 test set. The 1000 training images were augmented to 12000 Image with 6 frequencies of 2 main methods.

**Deep Network Architecture**

The Deep-Fill architecture consists of an encoder and a decoder part was presented in Figure 2 Encoder part of network takes an image and generates a high-dimensional feature vector, and encoder aggregate features at multiple levels. Decoder part of network takes a high dimensional feature vector and generates a ground truth like image, and decode features aggregated by encoder at multiple levels. The Deep-Fill network is composed of convolution operation, max pooling, ReLU activation, concatenation, and up convolution layers. The gap image is go through the network along with all possible paths (contraction, expansion, and concatenation paths) then gap free image were generated by decoder part of network.

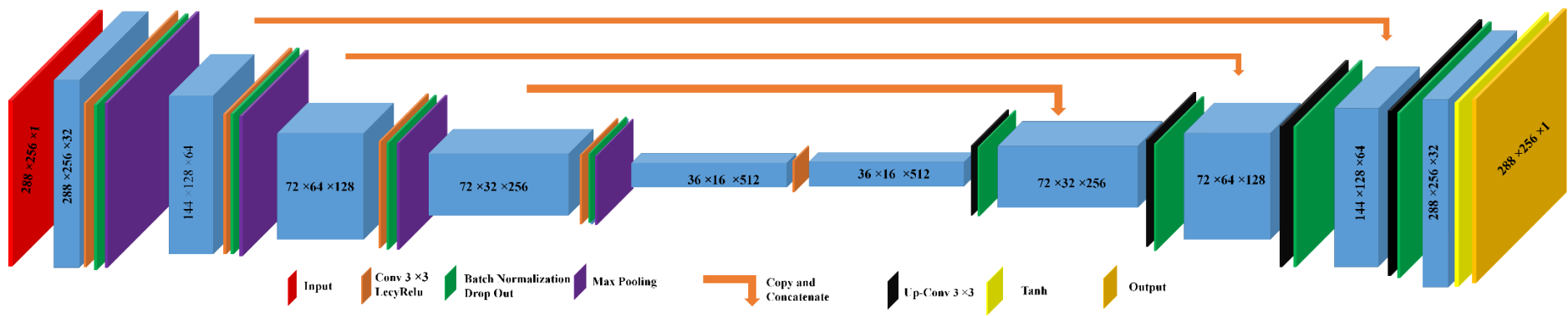

Figure 2: The main architecture of deep convolutional encoder decoder (Deep-Fill)

**Analysis**

Image quality of the generated images were quantitatively assessed by different quality metrics including: root mean squared error (RMSE), peak signal-to-noise ratio (PSNR), and structural Similarity index metrics (SSIM) and pixel wise correlation by using Pearson correlation coefficient ($R^2$).

For equations (1) and (2) N and M are the number of pixels in horizontal and vertical direction respectively. $GAPF(i,j)$ and $DeepFill(i,j)$ refer to the i and j coordinates of the gap free image and the synthesis image produced by $DeepFill$.

1) Root Mean Squared Error (RMSE): RMSE is square root or quadratic mean of differences between the AC images produced by Deep-Fill and GAPF and is defined as follows.

$$\text{RMSE} = \sqrt{\frac{1}{NM} \sum_i^N \sum_j^M (GAPF(i,j) - \text{Deep-Fill}(i,j))^2} \qquad \text{Eq. 1}$$

2) Peak Signal to Noise Ratio (PSNR): Ratio between the maximum possible power of a signal and noise where signal refers to the original image and noise refers to the standard error. PSNR calculated as follow.

$$\text{PSNR} = 10 \, Log_{10} \frac{(2^n-1)^2}{\sqrt{MSE}} \qquad \text{Eq. 2}$$

where n is maximum pixel value of the image.

3) Structural Similarity Index Metrics (SSIM)

SSIM measures the structural similarity between two images. SSIM is a perception-based measure that considers image degradation as perceived change in structural information. SSIM also incorporates important perceptual phenomena, including both luminance masking and contrast masking terms. SSIM where calculated by 4×4 kernel size as follow:

$$\text{SSIM}(x,y) = \frac{(2\mu_x\mu_y + c_1)(2\sigma_{xy} + c_2)}{(\mu_x^2 + \mu_y^2 + c_1)(\sigma_x^2 + \sigma_y^2 + c_2)} \qquad \text{Eq. 3}$$

where $\mu_x$ the average of Deep-Fill is, $\mu_y$ is the average of GAPF, $\sigma_x^2$ is variance of Deep-Fill, $\sigma_y^2$ is the variance of GAPF, $\sigma_{xy}$ is covariance of Deep-Fill and GAPF. Two variables $c_1$ and $c_2$ are used to stabilize the division with a weak denominator defined as:

$$c_1 = (K_1 L)^2, \ c_2 = (K_2 L)^2 \qquad \text{Eq. 4}$$

where L is the dynamic range of the pixel-values, $K_1$ and $K_2$ set by default to 0.01 and 0.03 respectively. Also, SSIM is a symmetric metric as it satisfies the condition of symmetry (i.e SSIM (Deep-Fill, GAPF) = SSIM (GAPF, Deep-Fill))**.** The values of SSIM range from 0 to 1 where a higher value indicates higher similarity.

**Result:**

Figure 3 to 5 shows the qualitative results of gap effect on reconstructed image quality against deep-fill based recovery method in three simulated brain phantoms. It can obviously be seen that deep-fill based method can significantly provide higher image quality than existent scanner images using FBP reconstruction method. We also added ideal/ground truth images in mentioned figures to have qualitative comparison between the results, surprisingly we cannot see the significant difference between recovered and ideal reconstructed images.

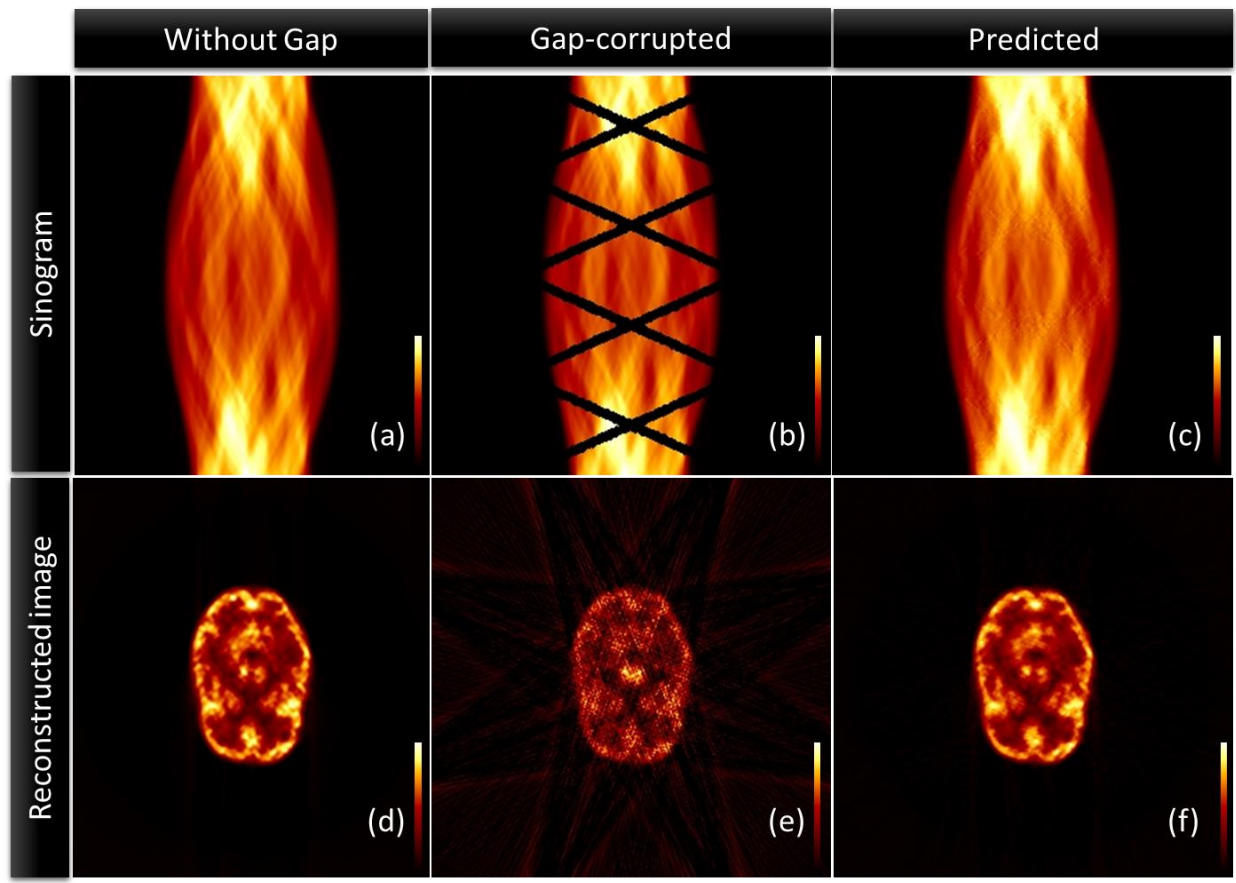

Figure 3: Gap filling results using deep recovery method for simulated Hoffman brain phantom. (a) Ideal sinogram without gap (b) Gap-corrupted sinogram (c) Predicted/recovered sinogram (d) Ideal reconstructed image without gap (e) Gap-corrupted reconstructed image (f) Predicted/recovered reconstructed image.

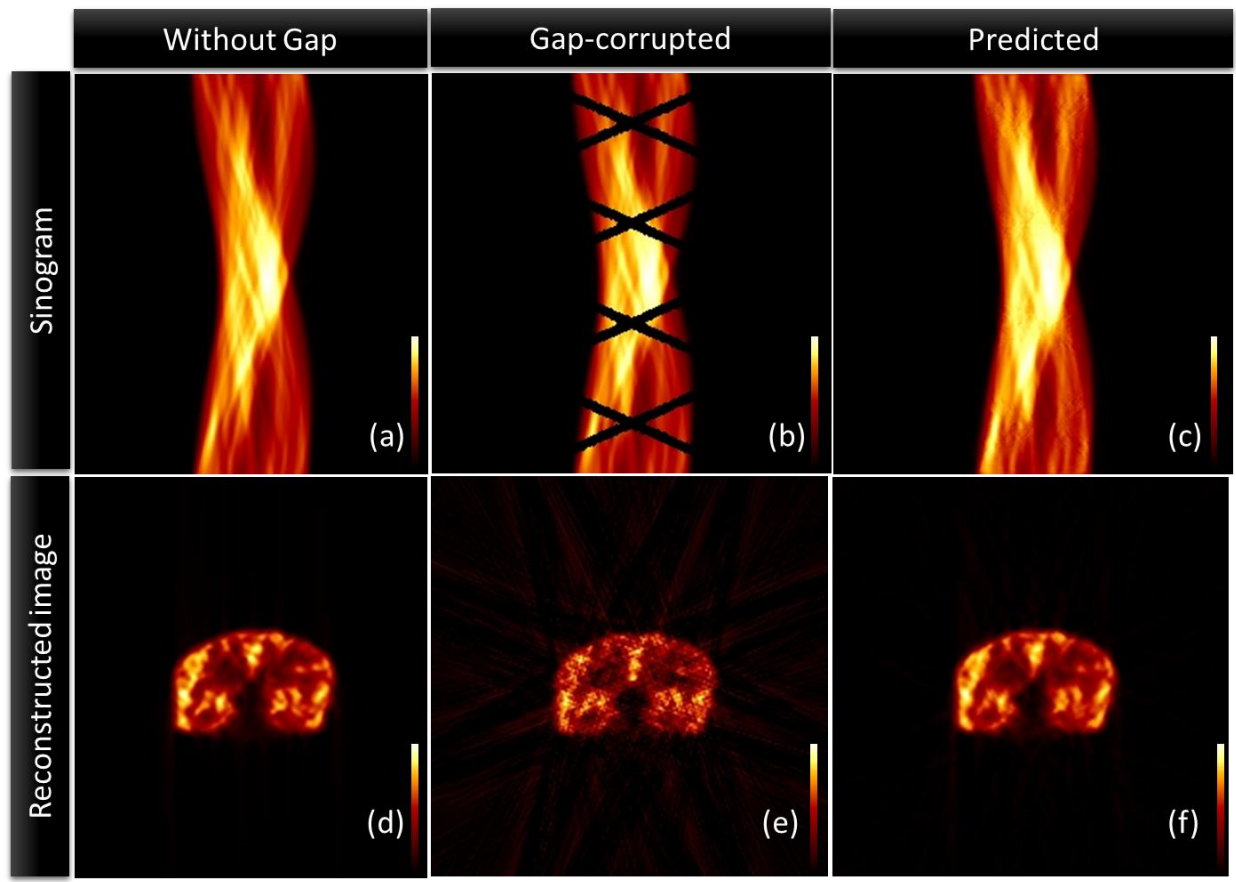

Figure 4: Gap filling results using deep recovery method for simulated XCAT phantom. (a) Ideal sinogram without gap (b) Gap-corrupted sinogram (c) Predicted/recovered sinogram (d) Ideal reconstructed image without gap (e) Gap-corrupted reconstructed image (f) Predicted/recovered reconstructed image.

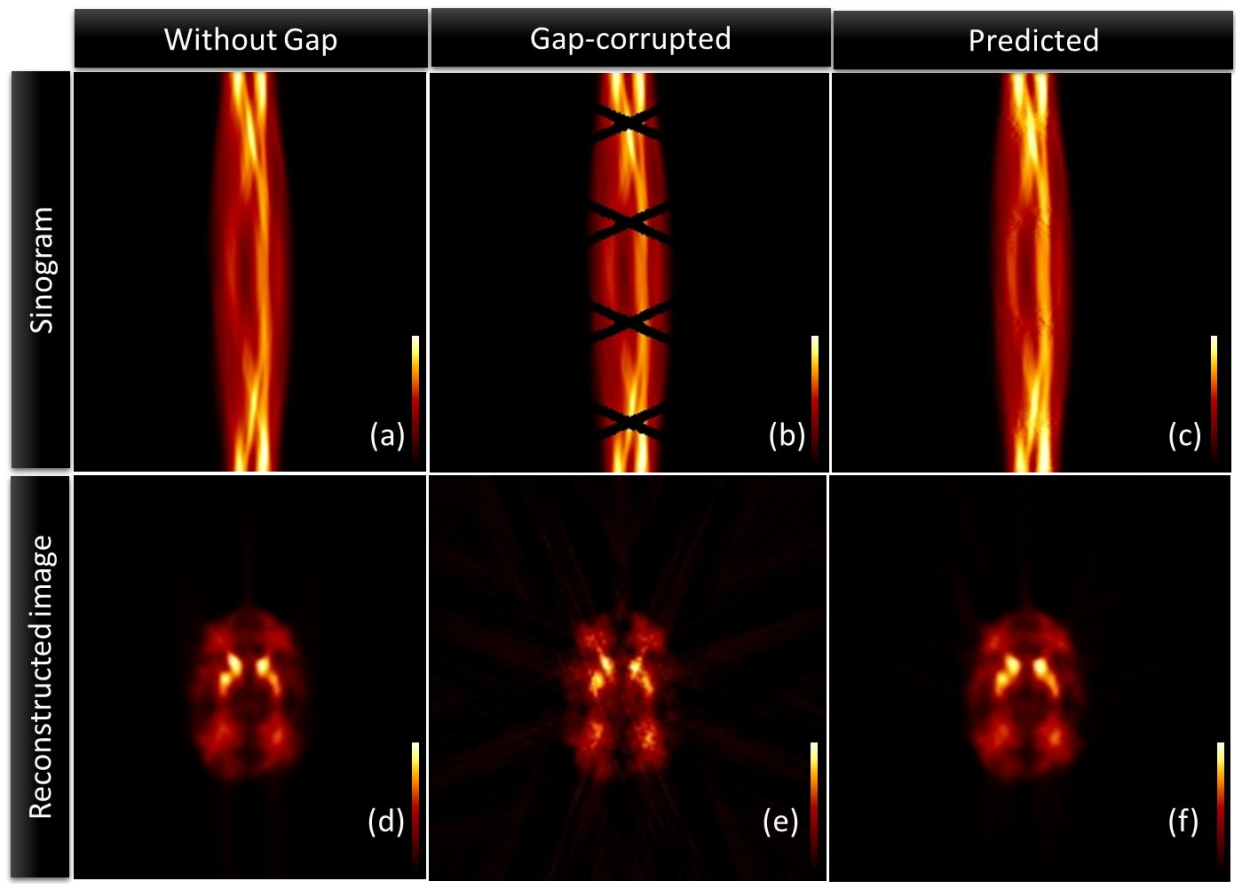

Figure 5: Gap filling results using deep recovery method for simulated Zubal brain phantom. (a) Ideal sinogram without gap (b) Gap-corrupted sinogram (c) Predicted/recovered sinogram (d) Ideal reconstructed image without gap (e) Gap-corrupted reconstructed image (f) Predicted/recovered reconstructed image.

Table 1 provides statistical description of root mean squared error (RMSE) between generated and ground truth images in different sets. The average RMSE was 0.00052±0.00034 in Sino-Test, Sino-Val and Recon-Test sets and 0.00058±0.00037 in Recon-Val set. The min and max RMSE were 0.00016 and 0.00134 in different sets. Figure 6 shows box plot of RMSE in these data sets.

Table 1: Statistical description of root mean squared error (RMSE) between generated and ground truth images in different data sets, Sino-Test; test Set of sinogram data, Sino-Val; external validation Set of sinogram data, Recon-Test; test set of reconstruction data, Recon-Val; external validation set of reconstruction data

| RMSE | Sino-Test | Sino-Val | Recon-Test | Recon-Val |
|---|---|---|---|---|
| mean | 0.00052 | 0.00052 | 0.00052 | 0.00058 |
| std | 0.00034 | 0.00034 | 0.00034 | 0.00037 |
| min | 0.00016 | 0.00016 | 0.00016 | 0.00020 |
| 25% | 0.00021 | 0.00021 | 0.00021 | 0.00025 |
| 50% | 0.00035 | 0.00035 | 0.00035 | 0.00042 |
| 75% | 0.00075 | 0.00076 | 0.00075 | 0.00083 |
| max | 0.00124 | 0.00124 | 0.00124 | 0.00134 |

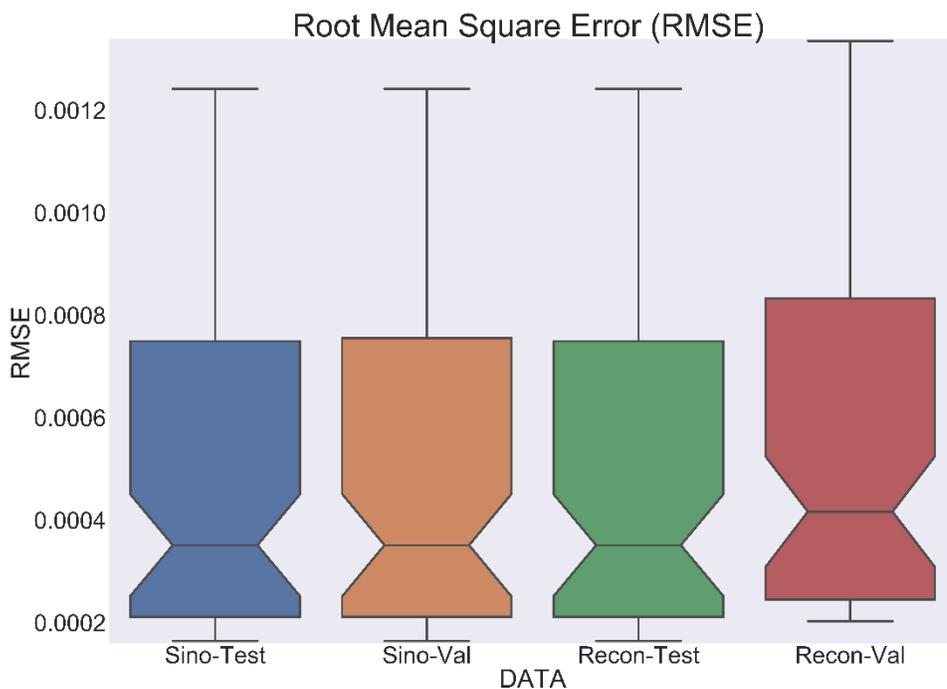

Figure 6: Box plot of root mean squared error (RMSE) in different data sets, Sino-Test; test Set of sinogram data, Sino-Val; external validation Set of sinogram data, Recon-Test; test set of reconstruction data, Recon-Val; external validation set of reconstruction data

Table 2 shows statistical description of peak signal-to-noise ratio (PSNR) between generated by Deep-Fill and ground truth images in different sets. The average PSNR was 33.82±3.01 in Sino-Test and Recon-Test sets and 33.51±3.01 in Sino-Val set. The minimum PSNR reported in Sino-Test and Recon-Test by 29.06 value and maximum PSNR reported in Sino-Test, Sino-Val and Recon-Test by 37.84 value. Figure 7 shows box plot of PSNR in these data sets.

Table 2: Statistical description of peak signal-to-noise ratio (PSNR) between generated and ground truth images in different data sets, Sino-Test; test Set of sinogram data, Sino-Val; external validation Set of sinogram data, Recon-Test; test set of reconstruction data, Recon-Val; external validation set of reconstruction data

| PSNR | Sino-Test | Sino-Val | Recon-Test | Recon-Val |
|---|---|---|---|---|
| mean | 33.82 | 33.51 | 33.82 | 33.32 |
| std | 3.01 | 3.01 | 3.01 | 2.93 |
| min | 29.06 | 29.22 | 29.06 | 28.74 |
| 25% | 31.25 | 30.76 | 31.25 | 30.79 |
| 50% | 34.54 | 32.05 | 34.54 | 33.80 |
| 75% | 36.76 | 36.60 | 36.76 | 36.10 |
| max | 37.84 | 37.84 | 37.84 | 36.92 |

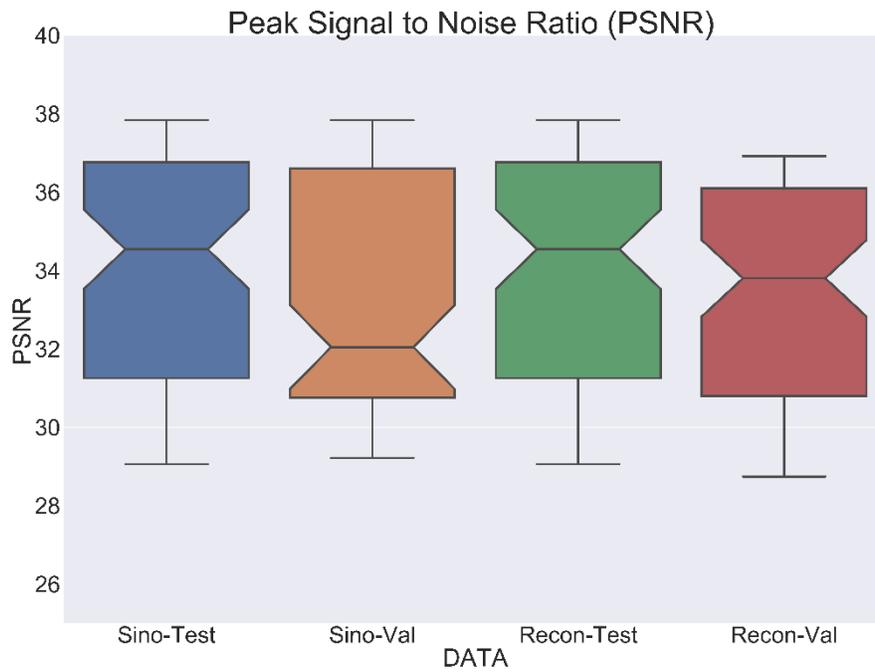

Figure 7: Box plot of peak signal-to-noise ratio (PSNR) in different data sets, Sino-Test; test Set of sinogram data, Sino-Val; external validation Set of sinogram data, Recon-Test; test set of reconstruction data, Recon-Val; external validation set of reconstruction data

Generated image similarity was assessed by using similarity index metrics (SSIM) presented in Table 3 followed by Figure 8. SSIM was higher than 0.99 in all dataset with standard deviation less than 0.001.

Table 3: Statistical description of structural Similarity index metrics (SSIM) between generated and ground truth images in different data sets, Sino-Test; test Set of sinogram data, Sino-Val; external validation Set of sinogram data, Recon-Test; test set of reconstruction data, Recon-Val; external validation set of reconstruction data.

| SSIM | Sino-Test | Sino-Val | Recon-Test | Recon-Val |
|---|---|---|---|---|
| mean | 0.999 | 0.991 | 0.999 | 0.999 |
| std | 0.001 | 0.008 | 0.001 | 0.001 |
| min | 0.996 | 0.974 | 0.996 | 0.995 |
| 25% | 0.999 | 0.982 | 0.999 | 0.998 |
| 50% | 0.999 | 0.997 | 0.999 | 1.000 |
| 75% | 0.999 | 0.998 | 1.000 | 1.000 |
| max | 0.999 | 0.999 | 0.999 | 0.999 |

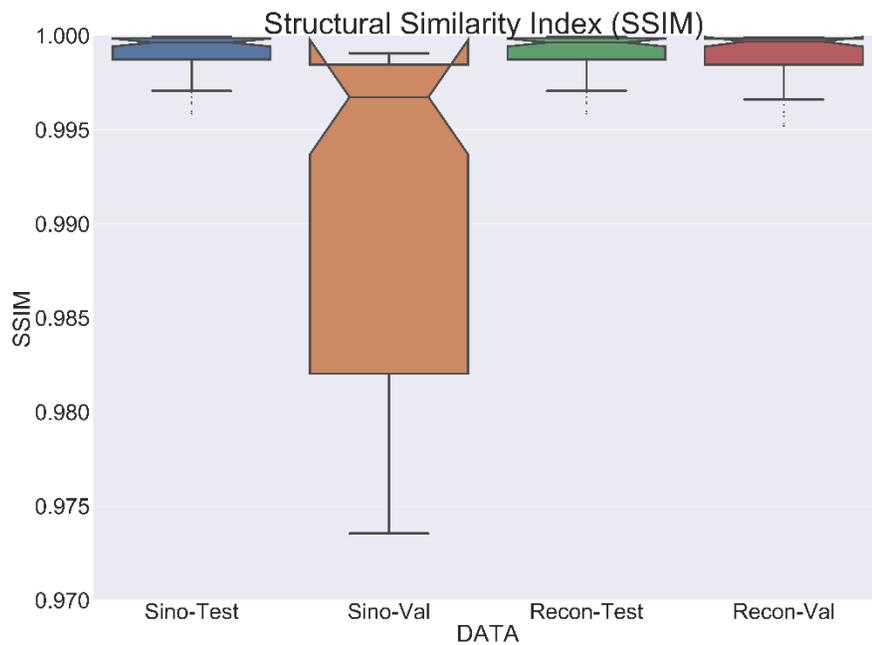

Figure 8: Box plot of structural Similarity index metrics (SSIM) in different data sets, Sino-Test; test Set of sinogram data, Sino-Val; external validation Set of sinogram data, Recon-Test; test set of reconstruction data, Recon-Val; external validation set of reconstruction data.

Table 4 depict statistical description of Pearson Correlation Coefficient ($R^2$) between generated and ground truth images in different sets. The average $R^2$ was higher than 0.96 in all with standard deviation less than 0.026. Figure 9 provides comparison of $R^2$ in schema of box plot in different sets.

Table 4: Statistical description of Pearson Correlation Coefficient between generated and ground truth images (PeCo, R2) in different data sets, Sino-Test; test Set of sinogram data, Sino-Val; external validation Set of sinogram data, Recon-Test; test set of reconstruction data, Recon-Val; external validation set of reconstruction data

| PeCo | Sino-Test | Sino-Val | Recon-Test | Recon-Val |
|---|---|---|---|---|
| mean | 0.971 | 0.966 | 0.971 | 0.964 |
| std | 0.026 | 0.024 | 0.026 | 0.031 |
| min | 0.918 | 0.918 | 0.918 | 0.906 |
| 25% | 0.954 | 0.954 | 0.954 | 0.940 |
| 50% | 0.960 | 0.958 | 0.960 | 0.960 |
| 75% | 0.996 | 0.996 | 0.996 | 0.994 |
| max | 0.997 | 0.997 | 0.997 | 0.996 |

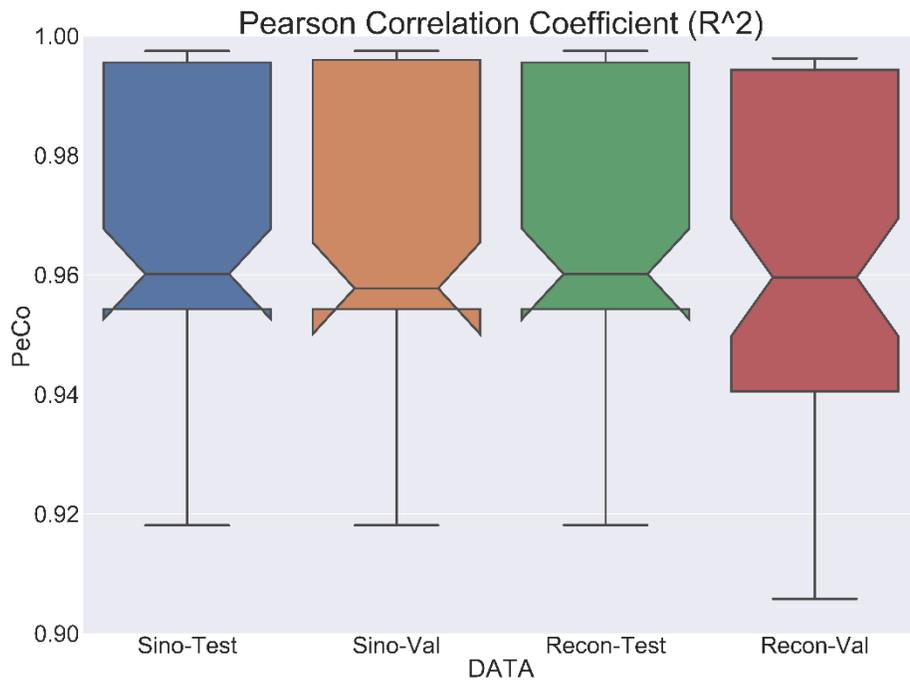

Figure 9: Box plot of Pearson Correlation Coefficient (PeCo, $R^2$) in different data sets, Sino-Test; test Set of sinogram data, Sino-Val; external validation Set of sinogram data, Recon-Test; test set of reconstruction data, Recon-Val; external validation set of reconstruction data

Figure 10 provide the pixel wise correlation between generated image using Deep-Fill and ground truth image of sinogram and reconstructed images in different data sets.

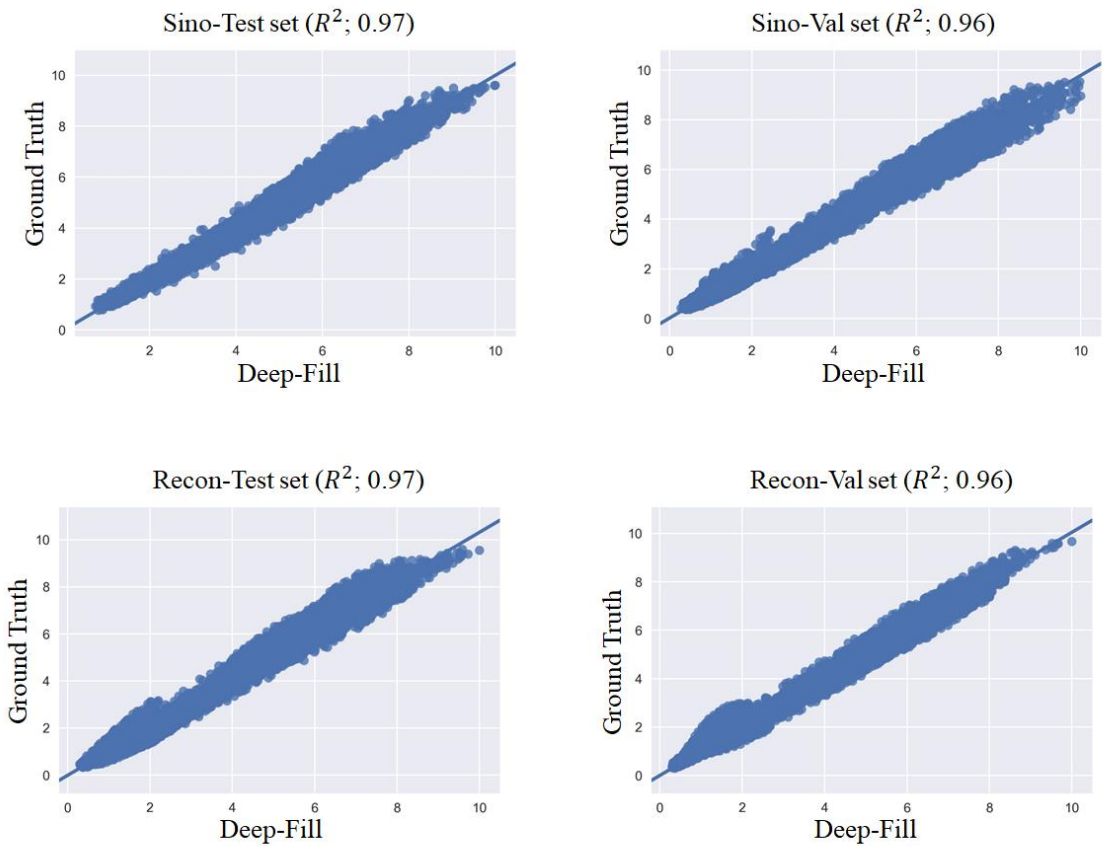

Figure 10: correlation of generated images by Deep-Fill and ground truth images using Pearson Correlation Coefficient (PeCo, $R^2$) in different data sets, Sino-Test; test Set of sinogram data, Sino-Val; external validation Set of sinogram data, Recon-Test; test set of reconstruction data, Recon-Val; external validation set of reconstruction data

**Discussion and Conclusion:**

In this study, we introduced and developed a novel method to recover missing data owing to inter detector gaps in PET scanners instrumentation. Our proposed method is based on deep learning approach that predict missing data in sinogram domain prior to reconstruction. For evaluation of the proposed method, multiple MC simulations using different voxel based brain phantoms were implemented. We compared quantitatively the results obtained from gap-corrupted sinograms of modeled HRRT scanner, sinograms without gap from ideal/ground truth HRRT scanner and predicted sinograms owing to the implemented our deep-fill method. Reconstructed images of each produced sinograms were analyzed as well.

According to the figure 3-5, it was obviously observed that gap-corrupted sinograms degraded the image quality, and deep-fill method could significantly predict sinogram and improve the image quality. Also the quantitatively obtained results in this study indicate the importance of applying a deep learning based approach as an appropriate gap-filling method. Our findings can be of specific importance not only in MC simulation studies, but also in practical real experiments, can play significant influence as the gap-filling procedure.

In this study, FBP algorithm was used for reconstruction because according the NEMA protocol, for measurement of the spatial resolution due to point sources must be reconstructed using FBP, therefore it is essential to introduce a robust FBP-based methods for the evaluation of spatial resolution for different scanners . FBP because of its fast operation with low computation load have many applications particularly in the dynamic kinetic studies and also is usually preferred in quantitative analysis of PET images because of its linear behavior(1, 7).

Note that although the proposed approach in this study was used for ECAT-HRRT PET scanner but is highly promising for different scanners and also for development of new PET designs with different configurations. Various PET designs with different detector configurations produce different gap patterns so for compensate the missing information owing to the gaps, deep learning–based approach could be a suitable potential method. For better demonstration of feasibility of deep-fill in gap filling in different PET scanner configurations, it is necessary to evaluate different design especially for small field of view scanners which are more sensitive to the produced gaps. Future work will examine the effect of deep-fill on different gap patterns owing to different scanner designs.

For neural network approaches especially for deep learning method, the most important limitation is the diversity and plenty of data used to train the model. In order to increase diversity in this study, we used three different simulated voxelized head phantoms with different uptake pattern. Also numerous slices of phantoms enhanced the data diversity and plenty. However, for robustness of deep-fill method outside of the head, for other organs or objects, it would be necessary to increase our train and test data.

In conclusion, in this study, we demonstrated that deep learning–based approaches applied to inter-detector gap filling can recover the missing data in sinogram and image and have the potential to improve the reconstructed image quality and to prevent degradation of PET image quantification. The flexibility of our proposed method is limited by our simulated data of HRRT brain scanner, but for future work, our promising idea must be evaluated for large gaps produced by various PET design with different simulated and realistic phantom and organ images.